\documentclass[iop]{emulateapj}
\usepackage{graphicx}
\usepackage{natbib}
\usepackage{aas_macros}
\usepackage{amsmath}
\usepackage{amssymb}
\usepackage{enumitem}
\usepackage[normalem]{ulem}

\newcommand{\lnu}{L_\mathrm{10-40\;keV}}
\newcommand{\Lnu}{$\lnu$}
\newcommand{\NH}{$N_\mathrm{H}$}
\newcommand{\nh}{N_\mathrm{H}}
\newcommand{\fcthick}{f_\mathrm{CThick}}
\newcommand{\Fcthick}{$\fcthick$}
\newcommand{\giv}{\;|\;}
\newcommand{\cmsq}{cm$^{-2}$}
\newcommand{\ergs}{erg~s$^{-1}$}

\newcommand{\refone}[1]{{#1}}

\shorttitle{First direct measurements of the $>$10 kev XLF for AGNs at z$>$0.1}
\shortauthors{Aird et al.}
\slugcomment{Accepted for publication in \apj}
\begin{document}

\title{The NuSTAR Extragalactic Survey: First direct measurements of the $\gtrsim$10 keV X-ray luminosity function for Active Galactic Nuclei at z$>$0.1}
\author{J.~Aird\altaffilmark{1,2}}
\author{D.~M.~Alexander\altaffilmark{2}} 
\author{D.~R.~Ballantyne\altaffilmark{3}}
\author{F.~Civano\altaffilmark{4,5,6}}
\author{A.~Del-Moro\altaffilmark{2}}
\author{R.~C.~Hickox\altaffilmark{6}} 
\author{G.~B.~Lansbury\altaffilmark{2}}
\author{J.~R.~Mullaney\altaffilmark{7}}
\author{F.~E.~Bauer\altaffilmark{8,9,10}}
\author{W.~N.~Brandt\altaffilmark{11,12,13}}
\author{A.~Comastri\altaffilmark{14}}
\author{A.~C.~Fabian\altaffilmark{1}}
\author{P.~Gandhi\altaffilmark{15,2}}
\author{F.~A.~Harrison\altaffilmark{16}}
\author{B.~Luo\altaffilmark{11,12}}
\author{D.~Stern\altaffilmark{17}}
\author{E.~Treister\altaffilmark{18}}
\author{L.~Zappacosta\altaffilmark{19}}
\author{M.~Ajello\altaffilmark{20}}
\author{R.~Assef\altaffilmark{21}}
\author{M.~Balokovi\'{c}\altaffilmark{16}}
\author{S.~E.~Boggs\altaffilmark{20}}
\author{M.~Brightman\altaffilmark{16}}
\author{F.~E.~Christensen\altaffilmark{22}}
\author{W.~W.~Craig\altaffilmark{20}} 
\author{M.~Elvis\altaffilmark{5}}
\author{K.~Forster\altaffilmark{16}}
\author{B.~W.~Grefenstette\altaffilmark{16}}
\author{C.~J.~Hailey\altaffilmark{23}}
\author{M.~Koss\altaffilmark{24}}
\author{S.~M.~LaMassa\altaffilmark{4}}
\author{K.~K.~Madsen\altaffilmark{16}}
\author{S.~Puccetti\altaffilmark{25,26}}
\author{C.~Saez\altaffilmark{27}}
\author{C.~M.~Urry\altaffilmark{4}}
\author{D.~R.~Wik\altaffilmark{28,29}}
\author{W.~Zhang\altaffilmark{30}}

\email{jaird@ast.cam.ac.uk}
\altaffiltext{1}{Institute of Astronomy, University of Cambridge, Madingley Road, Cambridge, CB3 0HA, U.K.}
\altaffiltext{2}{Centre of Extragalactic Astronomy, Department of Physics, Durham University, Durham, DH1 3LE, U.K.}
\altaffiltext{3}{Center for Relativistic Astrophysics, School of Physics, Georgia Institute of Technology, Atlanta, GA 30332, USA}
\altaffiltext{4}{Yale Center for Astronomy and Astrophysics, 260 Whitney Avenue, New Haven, CT 06520, USA}
\altaffiltext{5}{Harvard-Smithsonian Center for Astrophysics, 60 Garden St, Cambridge, MA 02138, USA}
\altaffiltext{6}{Department of Physics and Astronomy, Dartmouth College, 6127 Wilder Laboratory, Hanover, NH 03755, USA}
\altaffiltext{7}{The Department of Physics and Astronomy, The University of Sheffield, Hounsfield Road, Sheffield S3 7RH, U.K.}

\altaffiltext{8}{Instituto de Astrof\'{i}sica, Facultad de F\'{i}sica, Pontificia Universidad Cat\`{o}lica de Chile, 306, Santiago 22, Chile}
\altaffiltext{9}{Millennium Institute of Astrophysics, Santiago, Chile}
\altaffiltext{10}{Space Science Institute, 4750 Walnut Street, Suite 205, Boulder, Colorado 80301, USA}

\altaffiltext{11}{Department of Astronomy and Astrophysics, The Pennsylvania State University, 525 Davey Lab, University Park, PA 16802, USA}
\altaffiltext{12}{Institute for Gravitation and the Cosmos, The Pennsylvania State University, University Park, PA 16802, USA}
\altaffiltext{13}{Department of Physics, The Pennsylvania State University, University Park, PA 16802, USA}

\altaffiltext{14}{INAF Osservatorio Astronomico di Bologna, via Ranzani 1, I-40127, Bologna, Italy}

\altaffiltext{15}{School of Physics \& Astronomy, University of Southampton, Highfield, Southampton SO17 1BJ, U.K.}

\altaffiltext{16}{Cahill Center for Astrophysics, 1216 E. California Blvd, California Institute of Technology, Pasadena, CA 91125, USA}
\altaffiltext{17}{Jet Propulsion Laboratory, California Institute of Technology, 4800 Oak Grove Drive, Mail Stop 169-221, Pasadena, CA 91109, USA}

\altaffiltext{18}{Universidad de Concepci\`{o}n, Departamento de Astronom\'{i}a, Casilla 160-C, Concepci\`{o}n, Chile}
\altaffiltext{19}{Osservatorio Astronomico di Roma (INAF), via Frascati 33,
00040 Monte Porzio Catone (Roma), Italy}

\altaffiltext{20}{Space Sciences Laboratory, 7 Gauss Way, University of California, Berkeley, CA 94720-7450, USA}
\altaffiltext{21}{N\'{u}cleo de Astronom\'{i}a de la Facultad de Ingenier\'{i}a, Universidad Diego Portales, Av. Ej\'{e}rcito Libertador 441, Santiago, Chile}

\altaffiltext{22}{DTU Space, National Space Institute, Technical University
of Denmark, Elektrovej 327, DK-2800 Lyngby, Denmark}

\altaffiltext{23}{Columbia Astrophysics Laboratory, Columbia University, New York, NY 10027, USA}
\altaffiltext{24}{Institute for Astronomy, Department of Physics, ETH Zurich, Wolfgang-Pauli-Strasse 27, CH-8093 Zurich, Switzerland}
\altaffiltext{25}{ASDC-ASI, Via del Politecnico, I-00133 Roma, Italy}
\altaffiltext{26}{Osservatorio Astronomico di Roma (INAF), via Frascati 33,
00040 Monte Porzio Catone (Roma), Italy}
\altaffiltext{27}{Department of Astronomy, University of Maryland, College
Park, MD 20742-2421, USA}
\altaffiltext{28}{NASA Goddard Space Flight Center, Code 662, Greenbelt, MD 20771, USA}
\altaffiltext{29}{The Johns Hopkins University, Homewood Campus, Baltimore, MD 21218, USA}
\altaffiltext{30}{Physics \& Engineering Department, West Virginia Wesleyan College, Buckhannon, WV 26201, USA}

\begin{abstract}
We present the first direct measurements of the rest-frame 10--40 keV X-ray luminosity function (XLF) of Active Galactic Nuclei (AGNs) based on a sample of 94 sources at $0.1<z<3$, selected at 8--24~keV energies from sources in the \emph{NuSTAR} extragalactic survey program.
Our results \refone{are consistent with} the strong evolution of the AGN population 
seen in prior, lower-energy studies of the XLF. 
However, different models of the intrinsic distribution of absorption, which are used to correct for selection biases, give significantly different predictions for the total number of sources in our sample, leading to small, systematic differences in our binned estimates of the XLF.
Adopting a model with a lower intrinsic fraction of Compton-thick sources 
and a larger population of sources with column densities $\nh\sim10^{23-24}$~cm$^{-2}$ 
or a model with a stronger Compton reflection component (with a relative normalization of $R\sim2$ at all luminosities) can bring extrapolations of the XLF from 2--10~keV into agreement with our \emph{NuSTAR} sample.
Ultimately, X-ray spectral analysis of the \emph{NuSTAR} sources is required to break this degeneracy between the distribution of absorbing column densities and the strength of the Compton reflection component and thus refine our measurements of the XLF.
\refone{Furthermore, the models that successfully describe the high-redshift population seen by \emph{NuSTAR} tend to over-predict previous, high-energy measurements of the local XLF, indicating that there is evolution of the AGN population that is not fully captured by the current models.}

\end{abstract}

\maketitle

\section{Introduction}

Measurements of the luminosity function of Active Galactic Nuclei (AGNs) provide the key observational data available to track the history and distribution of accretion onto supermassive black holes.
X-ray surveys have been crucial for performing such measurements as they can efficiently identify AGNs down to low luminosities, over a wide range of redshifts, and where the central regions are obscured by large amounts of gas and dust \citep[see][for a recent review]{Brandt15}. 

A number of studies have presented measurements of the X-ray luminosity function (XLF) of AGNs based on surveys with the \textit{Chandra} or \textit{XMM-Newton} X-ray observatories \citep[e.g.][]{Ueda03,Barger05,Hasinger05,Aird10,Miyaji15}.
These studies show that the AGN population has evolved substantially over cosmic time. 
Both the space density of AGNs and the overall accretion density (which traces the total rate of black hole growth) peaked at $z\sim1-2$ and have declined ever since. 
The evolution also has a strong luminosity dependence, whereby the space density of luminous AGNs peaks at $z\approx2$ whereas lower luminosity AGNs peak at later cosmic times ($z\approx1$). 
This pattern, reflected in the evolution of the shape of the XLF, provides crucial insights into the underlying distributions of black hole masses and accretion rates \citep[e.g.][]{Aird13,Shankar13}.

A major issue in XLF studies with \textit{Chandra} and \textit{XMM-Newton} is the impact of absorption.
It is now well established that many AGNs are surrounded by gas and dust that obscures their emission at certain wavelengths \citep[e.g.][]{Martinez-Sansigre05,Stern05,Tozzi06}.
Soft X-rays ($\sim$~0.5--2 keV) will be absorbed by gas with equivalent neutral hydrogen column densities $\nh\gtrsim10^{22}$ \cmsq, whereas higher energy X-rays can penetrate larger column densities. 
Thus, samples selected at $\sim$~2--10 keV energies are less biased and are typically adopted in XLF studies \citep[e.g.][]{Aird10,Miyaji15}.  
However, absorption can still suppress the observable flux, especially at higher column densities ($\nh\gtrsim10^{23}$ \cmsq) and at lower redshifts ($z\lesssim1$), where the observed band probes lower rest-frame energies. 
The emission from Compton-thick sources ($\nh\gtrsim10^{24}$ \cmsq) is even more strongly suppressed, although such sources may still be identified at $\sim$~0.5--10 keV energies by their scattered emission and signatures of reflection from the obscuring material \citep[e.g.][]{Georgantopoulos13,Brightman14}.
Multiwavelength studies can also identify the signatures of heavy obscuration in X-ray detected AGNs \citep[e.g.][]{Cappi06,Alexander08,Gilli10,Georgantopoulos11} or directly identify additional, obscured AGN populations that are not detected in the X-ray band \citep[e.g.][]{Donley08,Juneau11,Eisenhardt12,DelMoro15}.

Correcting for absorption to accurately recover the XLF is challenging.
Luminosity estimates for individual sources can be corrected based on X-ray spectral analysis or hardness ratios \citep[e.g.][]{Ueda03,LaFranca05,Buchner15}.
Alternatively, less-direct statistical approaches can be adopted to constrain the underlying distribution of \NH\ and account for the impact on observed X-ray samples (e.g. \citealt{Miyaji15}; \citealt{Aird15}, hereafter A15). 
Despite much progress, the distribution of \NH\ as a function of luminosity and redshift (hereafter, ``the \NH\ function") and, most crucially, the fraction of Compton-thick sources remain poorly constrained outside the local Universe.

The cosmic X-ray background (CXB) provides an additional, integral constraint on the fraction of heavily obscured and Compton-thick AGNs.
At low energies ($\lesssim8$ keV), \textit{Chandra} surveys have resolved the majority ($\sim$~70--90\%) of the CXB into discrete point sources, predominantly unabsorbed and moderately absorbed AGNs at $z\sim0.5-2$ \citep[e.g.][]{Worsley05,Georgakakis08,Lehmer12,Xue12}.
However, the peak of the CXB occurs at much higher energies, $\sim$~20--30 keV \citep[e.g.][]{Marshall80,Churazov07,Ajello08}. 
Population synthesis models, based on studies of the XLF and \NH\ function at lower energies, attribute $\gtrsim70$\% of the emission at this peak to absorbed AGNs \citep[$\nh\gtrsim10^{22}$ \cmsq\ e.g.][]{Gilli07,Treister09,Draper09}, although the required fraction of Compton-thick ($\nh\gtrsim10^{24}$ \cmsq) AGNs is still uncertain \citep[e.g.][A15]{Ballantyne11,Akylas12}.
 
Until recently, only $\sim$1--2\% of the CXB at $\gtrsim10$ keV could be directly resolved into individual objects, due to the limited sensitivity at these energies achieved with non-focusing X-ray missions such as \emph{INTEGRAL} or \emph{Swift} \citep[e.g.][]{Krivonos07,Tueller08,Ajello12}. 
Nonetheless, AGN samples identified by these missions---which are less biased against obscured sources---have enabled crucial measurements of the XLF, the \NH\ function, and the fraction of Compton-thick AGNs in the local ($z\lesssim0.1$) universe \citep[e.g.][]{Beckmann09,Burlon11,Vasudevan13}, which are often used to extrapolate to higher redshifts \citep[e.g.][hereafter U14]{Ueda14}.

The \textit{Nuclear Spectroscopic Telescope Array} \citep[hereafter \emph{NuSTAR},][]{Harrison13} is the first orbiting observatory with $>10$ keV focusing optics, providing $\sim2$ orders of magnitude increase in sensitivity compared to previous high-energy observatories. 
One of the primary objectives of the \emph{NuSTAR} mission is to identify and characterize the source populations that produce the peak of the CXB. 
To this end, \emph{NuSTAR} is executing a multi-layered program of extragalactic surveys.
Source catalogs and initial results from the dedicated surveys of the COSMOS and ECDFS regions are presented by \citet[hereafter C15]{Civano15} and 
\citet[hereafter M15]{Mullaney15}, 
respectively, while \citet{Alexander13} presented the first results from our ongoing serendipitous survey program.
\citet{Harrison15} present source number counts at 3--8~keV and 8--24~keV energies from the full survey program and show that \emph{NuSTAR} is directly resolving $\sim35$\% of the CXB emission at 8--24~keV, a factor $\sim15-30$ times more than previous high-energy X-ray observatories. 

In this paper, we present the first measurements of the XLF of AGNs at $0.1<z<3$ based on direct selection of sources at hard ($>8$ keV) energies from across the \emph{NuSTAR} extragalactic survey program.
Section~\ref{sec:data} describes our data and defines our sample. 
Section~\ref{sec:method} describes our statistical methods to estimate intrinsic luminosities and recover the XLF. 
In Section~\ref{sec:xlf}, we present our measurements of the 10--40~keV XLF and explore the effects of different model assumptions.
We discuss our results and future prospects for the \emph{NuSTAR} survey program in Section~\ref{sec:discuss} and summarize our findings in Section~\ref{sec:conclusions}.
We adopt a flat cosmology with $\Omega_\Lambda = 0.7$ and $H_0 = 70$~km~s$^{-1}$~Mpc$^{-1}$  throughout this paper. 
\section{Data and sample selection}
\label{sec:data}

The \emph{NuSTAR} extragalactic survey program \citep[see][for an overview]{Harrison15} consists of three components: 
1) a deep ($\sim400$ ks) survey covering both the Extended Chandra Deep Field South (ECDFS: M15) and Extended Groth Strip (EGS: Aird et al., in preparation) regions; 
2) a medium depth ($\sim 100$ ks) survey covering the COSMOS field (C15);
and 3) a wide-area program searching for serendipitous detections across all \emph{NuSTAR} observations \citep[Fuentes et al., in preparation, Lansbury et al., in preparation]{Alexander13}. 
In this paper we select sources from across the \emph{NuSTAR} extragalactic survey program that are directly detected at 8--24 keV energies. 
\emph{NuSTAR} provides unprecedented sensitivity at these energies, although the sensitivity of this band is dominated by $\lesssim 12$~keV energies due to a combination of decreasing effective area and the decrease in source photon flux with increasing energy.

\begin{figure}
\includegraphics[width=\columnwidth,trim=30 10 0 0]{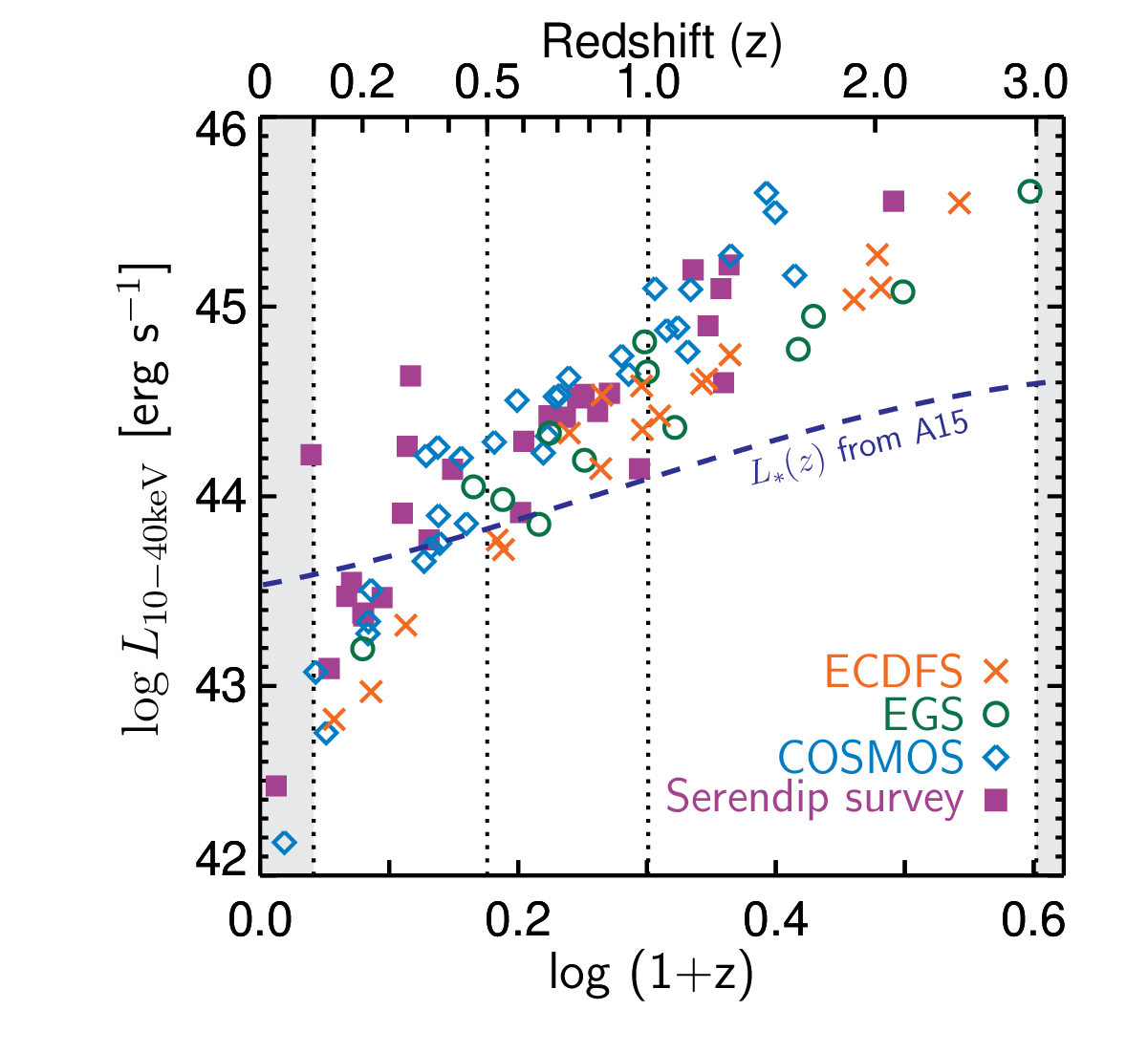}
\caption{
Rest-frame 10--40 keV X-ray luminosity (not corrected for absorption) versus redshift for sources in our 8--24 keV selected sample from the various \emph{NuSTAR} survey components, as indicated. 
Luminosities are estimated from the 8--24~keV observed fluxes (see Sections \ref{sec:data} and \ref{sec:luminosities} for details).
The vertical dotted lines show the limits of the redshift bins adopted for our XLF estimates in Section \ref{sec:xlf} (sources in the shaded regions are excluded). 
The blue dashed line indicates the characteristic break in the XLF, $L_*$, based on the 2--10 keV XLF measured by A15. 
}
\label{fig:lx_vs_z}
\end{figure}

Our overall sample consists of 97 sources. 
\refone{We identify lower energy X-ray counterparts to the vast majority of these sources (93 out of 97) and identify optical or infrared counterparts (matching to the low-energy X-ray position, if available) for all but one source (NuSTAR J033122-2743.9 in the ECDFS, discussed further below). 
Reliable (spectroscopic or photometric) redshift estimates are obtained for 91 out of our 97 sources.
Additional details are given in Sections \ref{sec:dedicated} and \ref{sec:serendip} below, with full details and catalogs provided by M15, C15 or Lansbury et al. (in prepartation).}
Figure~\ref{fig:lx_vs_z} shows the distribution of the rest-frame 10--40 keV luminosities versus redshift for our sample. 
Luminosities in this plot are estimated from the 8--24~keV count rates assuming an unabsorbed X-ray spectrum with photon index $\Gamma=1.9$ and a reflection component with a relative normalization of $R=1$, folded through the \emph{NuSTAR} response. Section~\ref{sec:luminosities} gives a detailed description of our spectral model and the uncertainties in these luminosity estimates, which are accounted for in our analysis of the XLF.
For our estimates of the XLF we use sources in the redshift range $0.1<z<3$, resulting in a sample of 94 sources (which includes an additional six sources with indeterminate redshifts). 

\subsection{Dedicated survey fields (ECDFS, EGS and COSMOS)}
\label{sec:dedicated}

In the dedicated survey fields (ECDFS, EGS and COSMOS), we adopt a consistent source detection procedure. 
We generate background maps using the \textsc{nuskybgd} code \citep{Wik14}, which we use to calculate the probability that the image counts in a $20\arcsec$ aperture were produced by a spurious fluctuation of the background (hereafter, the ``false probability") at the position of every pixel.
We then identify positions where the false probability falls below a set threshold and thus the observed counts can be associated with a real source.
See M15 and C15 for full details.

Our final sample of 8--24~keV detections consists of 19, 13 and 32 sources in the ECDFS, EGS and COSMOS fields, respectively.
We identify lower energy X-ray counterparts to our sources in the deep \emph{Chandra} or \emph{XMM-Newton} imaging of these fields \citep{Lehmer05,Xue11,Nandra15,Puccetti09,Cappelluti09} for all but one of our sources (NuSTAR J033122-2743.9 in the ECDFS, discussed by M15).
All of the low-energy counterparts have multiwavelength identifications, as given by M15 and C15 and references therein (for the ECDFS and COSMOS fields) or \citet{Nandra15} for the EGS field.
A very high fraction of these counterparts (86\%) have available spectroscopic redshifts; for the remainder we adopt the best photometric redshift estimate given by M15, C15 or \citet{Nandra15}. 
For two sources in the ECDFS (NuSTAR J033212-2752.3 and NuSTAR J033243-2738.3), which lack photometric estimates in the M15 catalog, we adopt photometric redshifts from \citet{Hsu14}.
We retain NuSTAR J033122-2743.9 (which lacks a low-energy or multiwavelength counterpart) in our sample but assume no knowledge of its redshift, effectively adopting a $p(z)$ distribution that is constant in $\log(1+z)$ (see A15 and further discussion in Section \ref{sec:luminosities} below). 
We note that this source could be a spurious detection (and would be consistent with the expected spurious fraction, given our false probability thresholds); this possibility is allowed for by our statistical methodology to determine the XLF.

\subsection{Serendipitous survey} 
\label{sec:serendip}

For the serendipitous survey, we adopt the same source detection procedure as in the dedicated fields but use a different method to determine the background maps since in many of the fields a bright target contaminates $\sim10-80$\% of the \emph{NuSTAR} field-of-view.
We thus take the original X-ray images and measure the counts in annular apertures of inner radius $30\arcsec$ and outer radius $90\arcsec$ centered at each pixel position.
We rescale the counts within each annulus to a $20\arcsec$ radius based on the ratio of the effective exposures.
This procedure produces maps giving estimates of the local background level at every pixel based on the observed images. The annular aperture ensures any contribution from a source at that pixel position is excluded from the background estimate. 
However, any large-scale contribution from a bright target source will be included.
We use these background maps, along with the mosaic counts images, to generate false-probability maps, and we proceed with source detection as in the dedicated survey fields, adopting a false-probability threshold of $<10^{-6}$ across all bands and fields. 
We exclude any detections within $90\arcsec$ of the target position.
We also exclude areas occupied by large, foreground galaxies (based on the optical imaging) or known sources that are associated with the target (but are not at the aimpoint).
In addition, we exclude areas where the effective exposure is $<30$\% of the maximal (on-axis) exposure in a given field, which removes unreliable detections close to the edge of the \emph{NuSTAR} field-of-view where the background is poorly determined.
We consider all fields analysed as part of the serendipitous program up to 2015 January 1, extending the sample of \citet{Alexander13}.
Full details of this extended serendipitous survey program will be given by Lansbury et al. (in preparation). 
For this paper, we apply a number of additional cuts:
\begin{enumerate}
\item We exclude fields at Galactic latitudes $\lvert b\rvert < 20\degr$, to ensure our sample is dominated by extragalactic sources.
\item We exclude any fields where there are $>10^{6}$ counts within $120\arcsec$ of the aimpoint; i.e. fields where the target is bright and will substantially contaminate the entire \emph{NuSTAR} field-of-view.
\item We only consider fields at declinations ${>-5\degr}$; i.e.~accessible from the Northern hemisphere.
\end{enumerate}

We do not expect any of these cuts to introduce systematic biases in the sample.
The final cut ensures that we have a high spectroscopic redshift completeness,\footnote{The cut on declinations $>-5\degr$ is not applied for the number counts analysis of \citet{Harrison15}, where redshift information is not required, resulting in a larger areal coverage and sample size.} thanks to a substantial ongoing follow-up program with Palomar and Keck (PI: Harrison; PI: Stern), in addition to existing redshifts from the literature. 
Follow-up programs of Southern fields are underway using Magellan (PIs: Bauer, Treister) and ESO NTT (PI: Lansbury) but have yet to achieve the high level of spectroscopic completeness required for the present study. 

After applying the above cuts, our serendipitious survey spans 106 \emph{NuSTAR} fields,
corresponding to a total area coverage of $4.40$ deg$^2$.
These \emph{NuSTAR} observations span a wide range of depths ($\sim10-1000$~ks, although predominantly $\lesssim50$~ks), resulting in a wide range of sensitivities (see Section \ref{sec:sens} below and Figure~\ref{fig:acurves}). 

Our serendipitous sample contains 33 sources that are detected in the 8--24~keV band. 
\refone{We identify lower energy counterparts to 30 of these sources, with the majority of counterparts (19/30) identified in the 3XMM source catalog \citep{Watson08,Rosen15} and the remainder identified manually in archival \textit{XMM-Newton}, \textit{Chandra}, or \textit{Swift}/XRT imaging data.
We also identify counterparts in the WISE all-sky survey \citep{Wright10} for all but one of our sources (including two of the sources that lack low-energy X-ray counterparts).
We identify optical counterparts, matching to the low-energy X-ray position or WISE positions (if available), using imaging from the SDSS \citep{York00}, USNOB1 \citep{Monet03}, or our own pre-imaging obtained during the spectroscopic  follow-up program. Full details on the cross-matching process for the full serendipitous survey sample will be provided by Lansbury et al. (in preparation).}

Out of our sample of 33 sources, 28 (85\%) have spectroscopic redshifts (all of these sources have an extragalactic origin). 
We retain the remaining five sources but assume no prior knowledge of their redshift, adopting a $p(z)$ distribution that is constant in $\log(1+z)$. 
However, after folding these broad $p(z)$ distributions through our models of the XLF a moderate redshift (mean $z\sim0.7$) is generally preferred. 
We note that $\sim3$\% of sources at $\lvert b \rvert >20\degr$ with spectroscopic classifications in our full serendipitous sample (including sources detected in the 3-8~keV and 3--24~keV bands) are associated with Galactic sources. Thus, one or more of our five 8--24~keV sources without spectroscopic classifications could have a Galactic origin; given our high overall redshift completeness, this potential contamination will have a negligible impact on our results.
A full list of fields and the properties of sources from the serendipitous survey program, marking the fields and sources used in this work, will be given by Lansbury et al. (in preparation).

\subsection{Sensitivity analysis}
\label{sec:sens}

Our source detection procedure (using false probability maps) is essentially identical across all the different survey components, allowing us to determine the sensitivity in a consistent manner. 
We construct sensitivity maps following the procedure described by \citet{Georgakakis08} that accounts for the Poisson nature of the detection.
First, for each pixel position in our images we estimate the minimum number of counts, $L$, in a 20\arcsec\ extraction aperture that would satisfy our false probability detection threshold given the local background estimate at that position, $B$. 
We then estimate the expected counts, $s$, from a source of flux, $F$, given the effective exposure and apply an aperture correction\footnote{As discussed by C15, the core of the \emph{NuSTAR} point spread function varies by less than a few percent over the field-of-view. Thus, we can neglect any spatial dependence of the aperture correction.} and fixed flux-to-count rate conversion factor from C15. 
We then calculate the probability that the combination of the expected source and background counts, $s+B$, produces a total number of counts that exceeds our threshold for detection, $L$. 
Each pixel contributes fractionally to the total area curve---the survey area sensitive to a given flux---in proportion to this probability.
We sum over all pixels to determine our overall area curves for a given survey component. 
Masked areas with low exposure or corresponding to the target (for the serendipitous survey fields) are excluded in this calculation. 

\begin{figure}
\includegraphics[width=\columnwidth,trim=20 0 0 0]{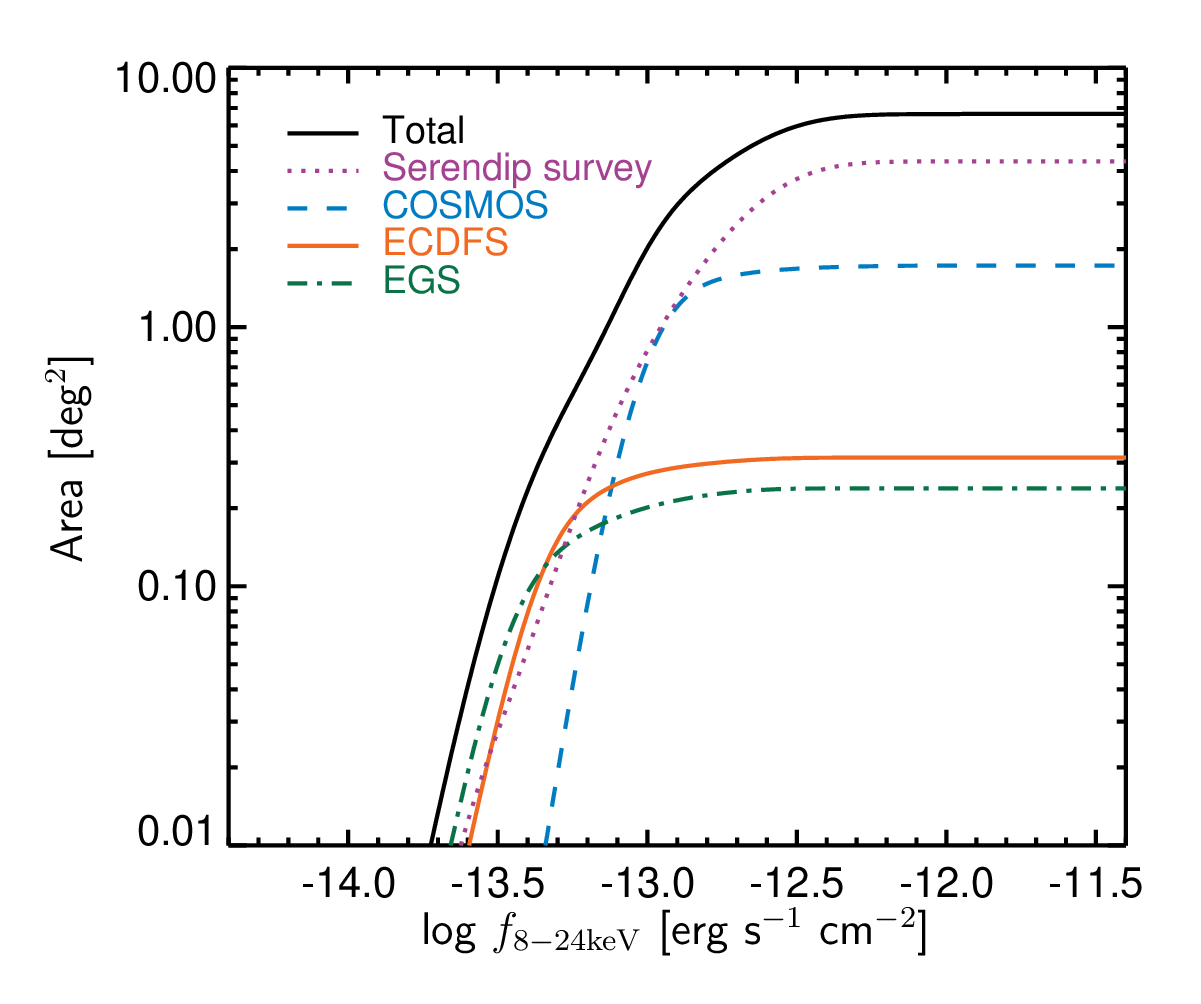}
\caption{X-ray area curves (area as a function of 8--24~keV flux) for the various \emph{NuSTAR} extragalactic survey components, as indicated.
}
\label{fig:acurves}
\end{figure}

We note that the true flux-to-counts conversion factor depends on the spectral shape of the source. 
In the XLF analysis described below (Section \ref{sec:method}), we allow for a range of spectral shapes when converting between luminosities and count rates, which are converted to an equivalent flux to determine the sensitivity from our area curves.
Thus, the dependence of the sensitivity on the spectral shape is accounted for in our analysis.

Figure~\ref{fig:acurves} shows the area coverage as a function of the 8--24~keV flux for the various survey components. 
Our COSMOS and ECDFS area curves are in good agreement with those derived from simulations by C15, verifying our analytic method. 
The serendipitous survey not only covers the largest overall area but also covers an area comparable to the dedicated deep surveys at fainter fluxes, making it a powerful addition to our study. 
We note that the dedicated surveys have very deep supporting data at lower X-ray energies and other wavelengths, which enables the high redshift completeness and will  be exploited in future studies of the X-ray and multiwavelength properties of \emph{NuSTAR} sources.

\section{Methodology}
\label{sec:method}

\subsection{Luminosity estimates} 
\label{sec:luminosities}

For each source in our sample, we must estimate the intrinsic luminosity, corrected for absorption and accounting for any other spectral features (e.g. reflection, scattering), allowing us to trace the accretion power of the AGNs in our sample.
We estimate the luminosity in the rest-frame 10--40 keV energy range, $L_\mathrm{10-40keV}$.
This luminosity must be corrected for the effects of absorption along the line-of-sight and includes the contribution of any reflection component.
The rest-frame 10--40 keV energy range roughly corresponds to the observed 8--24 keV \emph{NuSTAR} band (where we perform detection) for $z\approx0.3-1$ (where the bulk of our sample lies) and is becoming the standard for extragalactic \emph{NuSTAR} surveys \citep[e.g.][]{Alexander13,DelMoro14}.

To convert between the observed fluxes and $L_\mathrm{10-40keV}$ requires knowledge of the X-ray spectrum. 
X-ray spectral analysis of the \emph{NuSTAR} sources is underway \citep[see][Del Moro et al. in preparation and Zappacosta et al. in preparation]{Alexander13,DelMoro14}.
For this study, we adopt the statistical approach used by A15.
We use a particular X-ray spectral model: an absorbed power-law along with a simple modeling of the Compton reflection \citep[\emph{pexrav}:][]{Magdziarz95} and a soft scattered component.
Our model is described in detail in section 3.1 of A15. 
We use priors to describe the expected distributions of the spectral parameters.
For the photon index, $\Gamma$, we apply a normal prior with a mean of 1.9 and standard deviation of 0.2. 
For the scattered fraction, $f_\mathrm{scatt}$, we assume a lognormal prior with mean $\log f_\mathrm{scatt}=-1.73$ and a standard deviation of 0.8 dex.
For the reflection strength, $R$, we adopt a constant prior in the range $0<R<2$.
These priors allow for our lack of knowledge of the true values for an individual \emph{NuSTAR} source.
This simple spectral model should provide an adequate description of the \emph{broad} spectral properties of X-ray AGNs in the \emph{NuSTAR} bandpass. 
Based on this model, we can derive the joint probability distribution function for the intrinsic luminosity ($L_\mathrm{10-40keV}$), absorption column density (\NH), and redshift ($z$), for an individual source, $i$, in our sample, 
\begin{eqnarray}
p&&(\lnu,\nh, z \giv d_i, T_i, b_i) \propto \\
 && p(z \giv d_i) \; p(\lnu, \nh \giv z, T_i, b_i) \;\pi(\lnu, \nh, z) \nonumber
\label{eq:psource}
\end{eqnarray}
where $d_i$ represents any data on the redshift (e.g. optical spectroscopy or a photometric redshift) and $T_i$ and $b_i$ represent the \emph{NuSTAR} data: the total observed 8--24~keV counts in a $20\arcsec$ aperture, $T_i$, and the estimated background counts, $b_i$. The final term, $\pi(\lnu, \nh, z)$, represents any prior knowledge of the expected values of \Lnu, \NH\ and $z$.

In this study we assume $p(z \giv d_i)$ is given by a $\delta$-function at the available spectroscopic or photometric redshift value. 
We neglect the uncertainties in the photo-$z$ for the seven sources where we require such estimates, given their high accuracy \citep[$\sigma \lesssim 0.04$:][]{Luo10,Salvato11,Hsu14,Nandra15}.
For the six sources that lack redshift information completely (one source in ECDFS that lacks a \emph{Chandra} counterpart and five sources in the serendipitous survey without spectroscopic follow-up) we conservatively adopt a broad redshift distribution spanning $0<z<5$ that is a constant function of $\log(1+z)$.
Our high spectroscopic completeness (86\%) ensures that our XLF measurements are not significantly affected by errors in the photometric redshifts or sources with indeterminate redshifts. 

\begin{figure}
\begin{center}
\includegraphics[width=\columnwidth,trim=30 20 20 0]{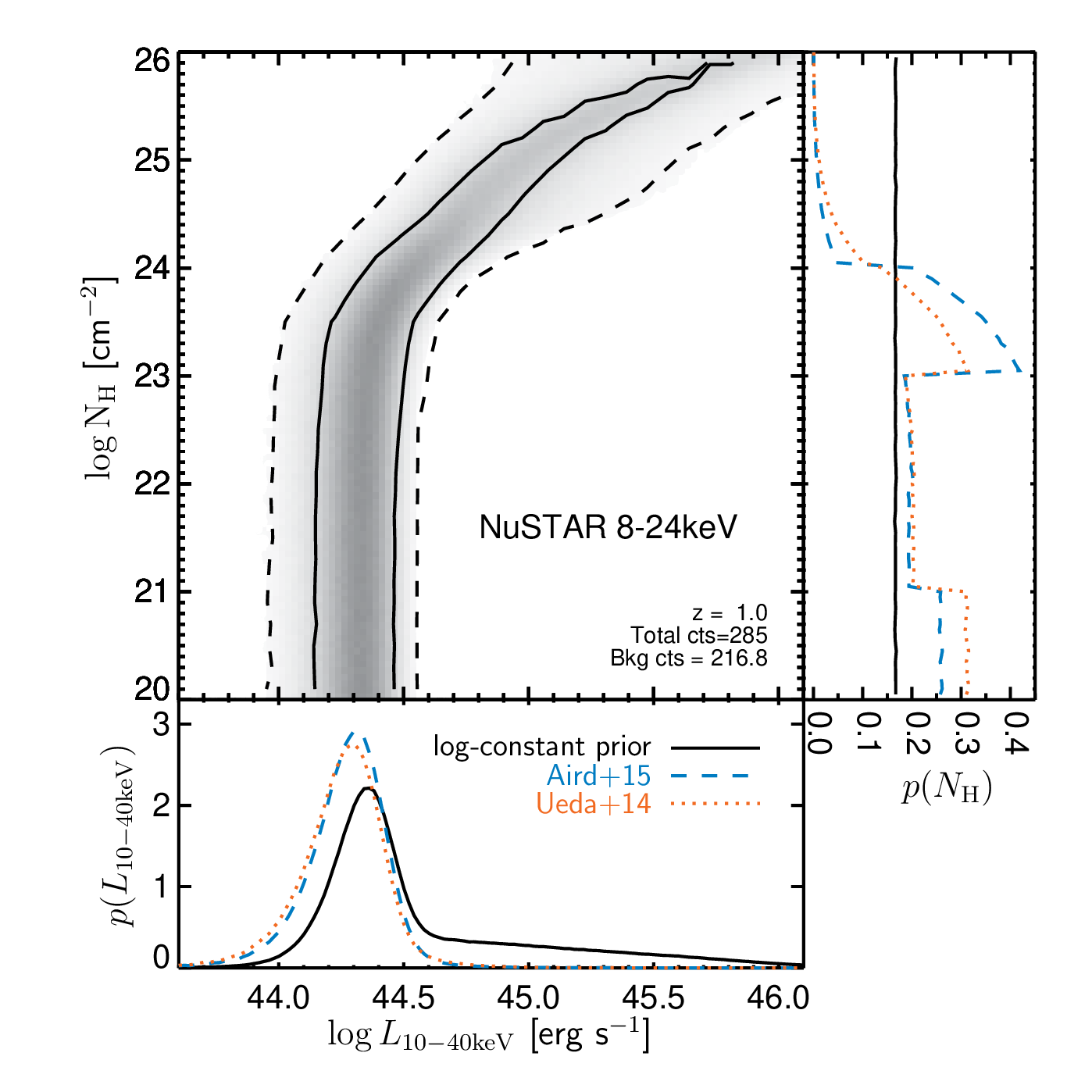}
\end{center}
\caption{Example of the two-dimensional probability distributions of \Lnu\ and \NH\ for a single  \emph{NuSTAR} 8-24 keV detection (\emph{main panel}), based purely on the observed 8-24 keV counts and background and assuming no \emph{a priori} knowledge of \Lnu\ or \NH. Contours indicate the 68.3\% (solid) and 95.4\% (dashed) confidence intervals (i.e. 1 and 2$\sigma$ equivalent) in the two-dimensional parameter space. 
The sub-panels show the marginalized distributions of \Lnu\ (\emph{bottom}) and \NH\ (\emph{right}) for the log-constant prior (black solid line) and applying informative priors based on previous estimates of the XLF and \NH\ function by A15 (blue dashed line) and U14 (red dotted line), extrapolated to the \emph{NuSTAR} energy band.
See Section \ref{sec:luminosities} for more details.
}
\label{fig:lx_vs_nh}
\end{figure}

We calculate $p(\lnu, \nh \giv z, T_i, b_i)$ by assuming that the observed counts are described by a Poisson process
and integrating over the priors on the spectral parameters ($\Gamma$, $f_\mathrm{scatt}$, and $R$), as described in section 3.1 of A15.
Our X-ray spectral model is folded through the \emph{NuSTAR} response function; thus we account for the variations in sensitivity across the 8--24 keV band.
The main panel of Figure~\ref{fig:lx_vs_nh} shows an example of the two-dimensional constraints on \Lnu\ and \NH\ for a typical source in our sample, assuming no \emph{a priori} knowledge of these values (adopting log-constant priors for \Lnu\ and \NH). 
The value of \NH\ is unconstrained, but the observed counts allow us to place constraints on the value of \Lnu, depending on the value of \NH.
If $\nh\lesssim10^{23}$ cm$^{-2}$ then, for this example, we estimate that $\log (\lnu /$\ergs$) \approx 44.35 \pm0.15$. 
The error is a combination of the Poisson uncertainties and the uncertainties on the other spectral parameters (primarily $\Gamma$ and $R$) but is not affected by the absorption.  
For higher values of \NH, the same observed counts must correspond to a higher intrinsic luminosity.

However, we do not completely lack \emph{a priori} knowledge of \Lnu\ or \NH. 
A number of previous studies have presented estimates of the XLF and \NH\ function of AGNs, albeit based on lower energy or lower redshift data \citep[e.g.][A15]{LaFranca05,Burlon11,Ueda14}.
Such studies tell us that high luminosity sources are significantly rarer than lower luminosity sources (due to the double power-law shape of the XLF) and predict different distributions of \NH. 
We can use these studies to apply informative priors when estimating $p(\lnu, \nh, z \giv d_i, T_i, b_i)$ for an individual source.

The sub-panels of Figure~\ref{fig:lx_vs_nh} illustrate the marginalized distributions of \Lnu\ and \NH\ (bottom and right panels respectively).
For our non-informative (log-constant) priors on \Lnu\ and \NH\ (solid black lines), there is a long tail in $p(\lnu)$, corresponding to Compton-thick column densities and correspondingly higher luminosities. 
We also apply priors based on the XLF and \NH\ functions of A15 (blue dashed lines, see section 5.1 and table 9 of A15 for the model specification and parameter values, respectively) and U14 (red dotted lines, see tables 2 and 4 of U14)
These studies give the XLF in terms of the intrinsic (i.e. absorption-corrected) rest-frame 2--10 keV luminosity, which we convert to \Lnu\ using our X-ray spectral model (marginalizing over the priors on the spectral parameters).
With priors based on either the U14 or A15 models, the long tail in $p(\lnu)$ is significantly suppressed as high luminosity sources are  substantially rarer (according to the XLF).
In addition, the peak of $p(\lnu)$ is shifted to slightly lower \Lnu\ than with a constant prior.
This shift accounts for the effect of Eddington bias: a detection is more likely to be a lower luminosity (and hence more common) source where the observed flux is a positive fluctuation. 
Our extensive simulations have shown that this effect is significant in our \emph{NuSTAR} survey data (see C15).

The distribution of $p(\nh)$ (right panel of Figure~\ref{fig:lx_vs_nh}) is mainly determined by the shape of the \NH\ function at $\nh\lesssim10^{23}$ \cmsq.
Thus, in this example, the probability density is slightly higher for $\nh=10^{20-21}$~\cmsq\  than for $\nh=10^{21-23}$~\cmsq.
There are only slight differences between the priors based on U14 or A15.
At a fixed luminosity, the intrinsic \NH\ function rises at $\nh=10^{23-24}$~\cmsq\ (for both models), hence $p(\nh)$ increases at $\nh=10^{23}$~\cmsq.
However, at $z=1.0$, absorption by column densities $\gtrsim10^{23}$~\cmsq\ starts to suppress the observed 8--24 keV flux; thus, at $\nh>10^{23}$~\cmsq\ the same observed counts must correspond to a higher \Lnu.
As higher luminosity sources are rarer, the marginalized $p(\nh)$ decreases as \NH\ increases, indicating that a detected source is less likely to be associated with these higher levels of absorption. 
An individual detection is very unlikely to be a Compton-thick AGN and the chance of extreme column densities ($\nh\gtrsim10^{25}$ \cmsq) is even more strongly suppressed. 
Nonetheless, the probability of an individual source having a Compton-thick \NH\ is slightly higher for the U14 prior (which has a higher intrinsic Compton-thick fraction\footnote{We define the intrinsic fraction of Compton-thick AGNs, $f_\mathrm{CThick}$, as the ratio of the number of sources with ${\nh=10^{24-26}}$~\cmsq\ to all absorbed ($\nh>10^{22}$~\cmsq) AGNs.}, $\fcthick\approx50$\%) than for the A15 prior ($\fcthick\approx25$\%).

\subsection{Binned estimates of the X-ray luminosity function}
\label{sec:binned}

Our \emph{NuSTAR} sample is relatively small and is limited to $\gtrsim L_*$ sources at $z>0.5$ (see Figure \ref{fig:lx_vs_z}).  
Thus, we do not attempt to fit an overall parametric model for the shape of the XLF and its evolution.
Instead, we produce binned estimates of the XLF in fixed luminosity and redshift bins.
We adopt the $N_\mathrm{obs}/N_\mathrm{mdl}$ method \citep[]{Miyaji01,Aird10,Miyaji15} and use parametric models from previous, lower-energy studies of the XLF and \NH\ function to correct for underlying selection effects.
The binned estimate of the XLF is given by
\begin{equation}
\phi_\mathrm{bin} \approx \phi_\mathrm{mdl}(L_b, z_b) \ \frac{N_\mathrm{obs}}{N_\mathrm{mdl}}
\label{eq:nobsnmdl}
\end{equation}
where $\phi_\mathrm{mdl}(L_b, z_b)$ is a given model estimate of the XLF, evaluated at $L_b$ and $z_b$\footnote{We fix $L_b$ at the center of the luminosity bin, whereas $z_b$ is fixed at the mean redshift of all sources in a given redshift bin.}, and is rescaled by the ratio of the \emph{effective} observed number of sources in a bin ($N_\mathrm{obs}$) to the predicted number based on the model  ($N_\mathrm{mdl}$). 

To calculate $N_\mathrm{mdl}$ we fold a given model of the XLF and \NH\ function (A15 or U14) through the 8--24~keV sensitivity curves for our surveys (see Section \ref{sec:sens}, Figure~\ref{fig:acurves}).
We convert between \Lnu, \NH, and $z$ and the observed 8-24 keV flux based on our X-ray spectral model and the appropriate \emph{NuSTAR} response function. 
The $N_\mathrm{mdl}$ term accounts for selection biases in our sample, primarily driven by the underlying \NH\ function, and allows for the fact that heavily absorbed sources are expected to be under-represented in our sample. 

To calculate $N_\mathrm{obs}$ we sum the individual $p(\lnu, \nh,z \giv d_i, T_i,b_i)$ distributions of our sources (including priors based on either the U14 or A15 models) and integrate over a given luminosity-redshift bin. 
As we allow for a range of possible luminosities (and in some cases redshifts), an individual source can make a partial contribution to $N_\mathrm{obs}$ for multiple bins. 

We estimate errors on our binned XLF based on the approximate Poisson error in the effective observed number of sources in a bin, $N_\mathrm{obs}$. 
We obtain Poisson errors based on \citet{Gehrels86}, although in many cases $N_\mathrm{obs}$ is non-integer and thus these errors are approximations \citep[see also][]{Aird10,Miyaji15}.
We only plot bins where $N_\mathrm{obs}\ge1$.

We note that all three terms in Equation \ref{eq:nobsnmdl} will depend, to varying extents, on the underlying parametric model of the XLF and \NH\ function that is assumed.
In particular, changing the assumed \Fcthick\ will alter our binned estimates; we expect Compton-thick sources are severely under-represented in our observed sample, which is accounted for in the $N_\mathrm{mdl}$ term, and thus our XLF estimates are increased to allow for this missed population.
We explore these effects further in Section \ref{sec:xlf} below.

\section{Measurements of the X-ray luminosity function with \emph{NuSTAR}}
\label{sec:xlf}

\begin{figure*}
\includegraphics[width=\textwidth,trim=10 10 0 0]{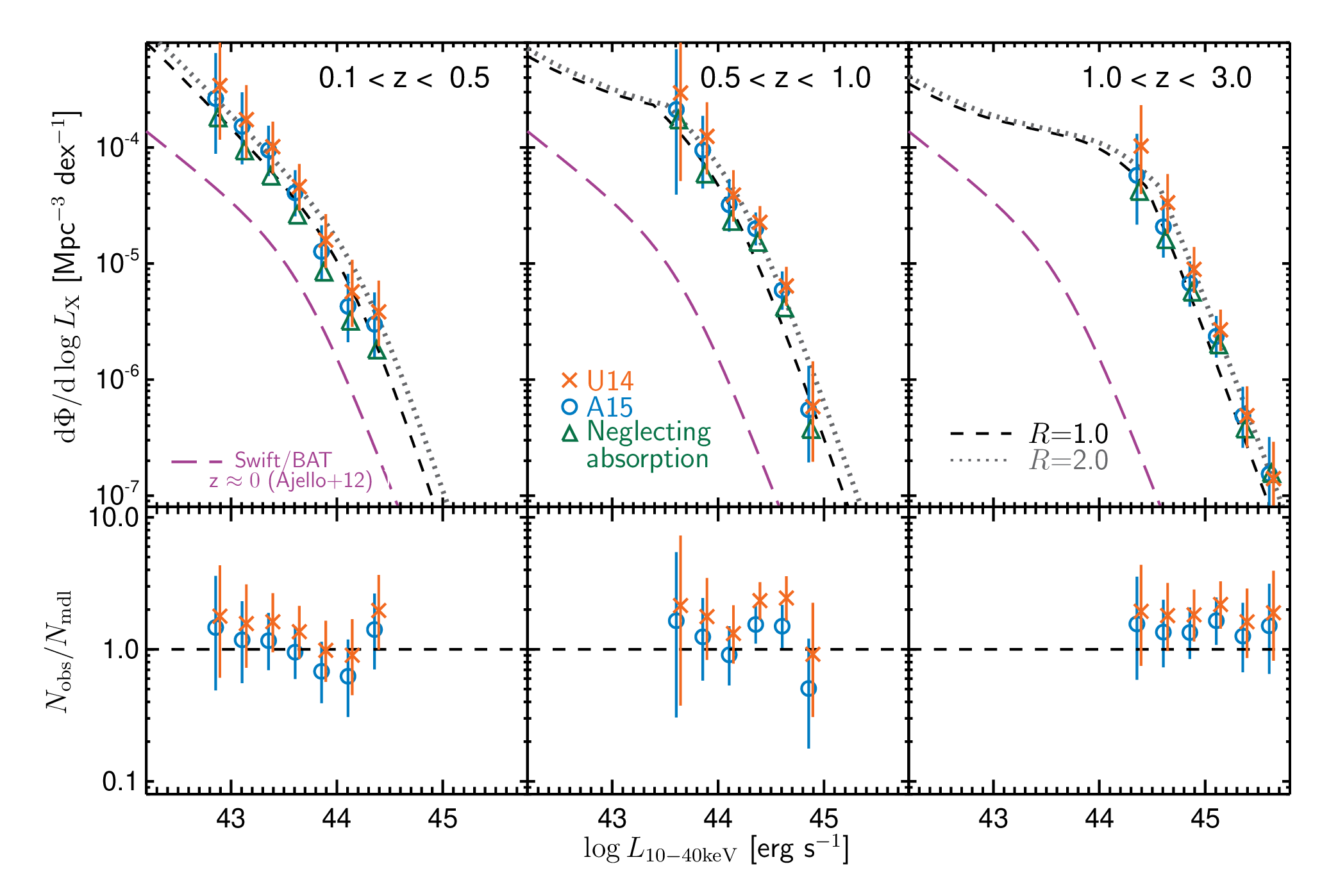}
\caption{
\emph{Top panels:}
Binned estimates of the 10--40~keV XLF of AGNs based on the \emph{NuSTAR} 8--24~keV selected sample in three redshift ranges.
Green triangles are estimates when the effects of absorption
are neglected (error bars are omitted for clarity).
The red crosses and blue circles are binned estimates of the XLF, where the luminosities 
\refone{and space densities} are corrected for absorption effects.
We account for absorption-dependent selection effects using the $N_\mathrm{obs}/N_\mathrm{mdl}$ method, assuming different models of the \NH\ function \refone{(including Compton-thick AGNs)}:
U14 and A15 for the red crosses and blue circles, respectively. 
Errors are based on the Poisson error in the observed number of sources in a bin.
There are small discrepancies between the U14 and A15 binned estimates due to differences in the models of the \NH\ function (see Figure \ref{fig:nhdist} and text in Section \ref{sec:xlf} for details).  
The black dashed line shows a model of the total XLF of AGNs (from U14, although the A15 XLF model is virtually identical over the range probed by our data), extrapolated from 2--10~keV energies to the 10--40~keV band assuming our baseline, unabsorbed X-ray spectral model (with $\Gamma=1.9$ and $R=1$). 
Assuming stronger reflection (e.g. $R=2$) shifts the model to higher luminosities (grey dotted line) and can thus bring the model into better agreement with the binned estimates based on the U14 \NH\ function.
The purple long-dashed lines in all panels show the XLF from \textit{Swift}/BAT \refone{ \citep{Ajello12}} at $z\approx0$, converted to the 10--40~keV band.
\emph{Bottom panels:}
Ratio between the XLF model and the binned estimates in terms of $N_\mathrm{obs}/N_\mathrm{mdl}$. 
}
\label{fig:xlfbins}
\end{figure*}

In this section we present measurements of the XLF of AGNs based on the \emph{NuSTAR} 8--24 keV sample.
In Figure~\ref{fig:xlfbins} (top panels) we present binned estimates of the 10--40 keV XLF in three redshift ranges based on the $N_\mathrm{obs}/N_\mathrm{mdl}$ method.
The green triangles are estimates where we neglect the effects of absorption, providing the most direct estimate of the \emph{observed} 10--40~keV XLF. 
We assume all sources have $\nh=10^{20}$~\cmsq\ and calculate the binned XLF utilising the best-fitting model for the Compton-thin XLF from A15, although these binned estimates are not strongly affected by the assumed XLF model.  
The red crosses and blue circles are binned estimates where we account for absorption effects using the best-fitting model of the \NH\ function from U14 and A15, respectively, and adopt the total XLF of AGNs (including Compton-thick sources), extraoplated from 2--10~keV energies.
These binned estimates are higher than when absorption is neglected (green triangles) due to the corrections for heavily absorbed ($\nh\gtrsim10^{23}$ \cmsq) and Compton-thick ($\nh\gtrsim10^{24}$ \cmsq) sources that will be under-represented in our observed samples due to selection biases, even at the harder energies probed by \emph{NuSTAR}.
Our binned estimates are listed in Table~\ref{tab:xlf}.

In Figure~\ref{fig:xlfbins}, we also show the U14 
model for the total XLF (including Compton-thick sources), extrapolated from 2--10 keV energies (black dashed line).
To extrapolate, we assume our (unabsorbed) X-ray spectral model evaluated at the mean of the priors on the spectral parameters (i.e. $\Gamma=1.9$, $f_\mathrm{scatt}=$1.9\%, and $R=1.0$), which gives 
\begin{equation}
\log (\lnu / \mathrm{erg\;s^{-1}}) \approx \log (L_\mathrm{2-10\;keV} / \mathrm{erg\;s^{-1}}) + 0.14
\label{eq:ref1}
\end{equation}
where $L_\mathrm{2-10keV}$ is the rest-frame 2--10~keV luminosity. 
Both the U14- and A15-based binned estimates, to first order, are in agreement with this extrapolation of the XLF model, within their errors.
We note that the extrapolation of the A15 model \refone{for the total XLF (including Compton-thick AGNs)} is virtually identical to the U14 XLF model over the range of luminosities and redshifts probed by our data and thus is also in agreement with the binned estimates.

On closer examination, however, there are differences between the binned estimates when the U14 model of the \NH\ function is used to correct for selection biases, rather than A15.
The U14 binned estimates are slightly higher than the A15 binned estimates, most noticeably at higher redshifts,
and are systematically higher than the extrapolated XLF model.
The $N_\mathrm{obs}/N_\mathrm{mdl}$ ratios, indicating the differences between our binned estimates and the model, are shown in the bottom panels of Figure~\ref{fig:xlfbins} and also illustrate this pattern: $N_\mathrm{obs}/N_\mathrm{mdl}$ is systematically higher for the U14-based binned estimates, indicating that the U14 model of the \NH\ function \emph{under}-predicts the number of sources in our \emph{NuSTAR} sample (i.e.~under-predicts $N_\mathrm{mdl}$).
These discrepancies must be due to differences in the underlying model of the \NH\ function, which introduces different corrections for the fraction of heavily absorbed and Compton-thick sources.

\begin{figure*}
\includegraphics[width=\textwidth, trim=10 10 0 0]{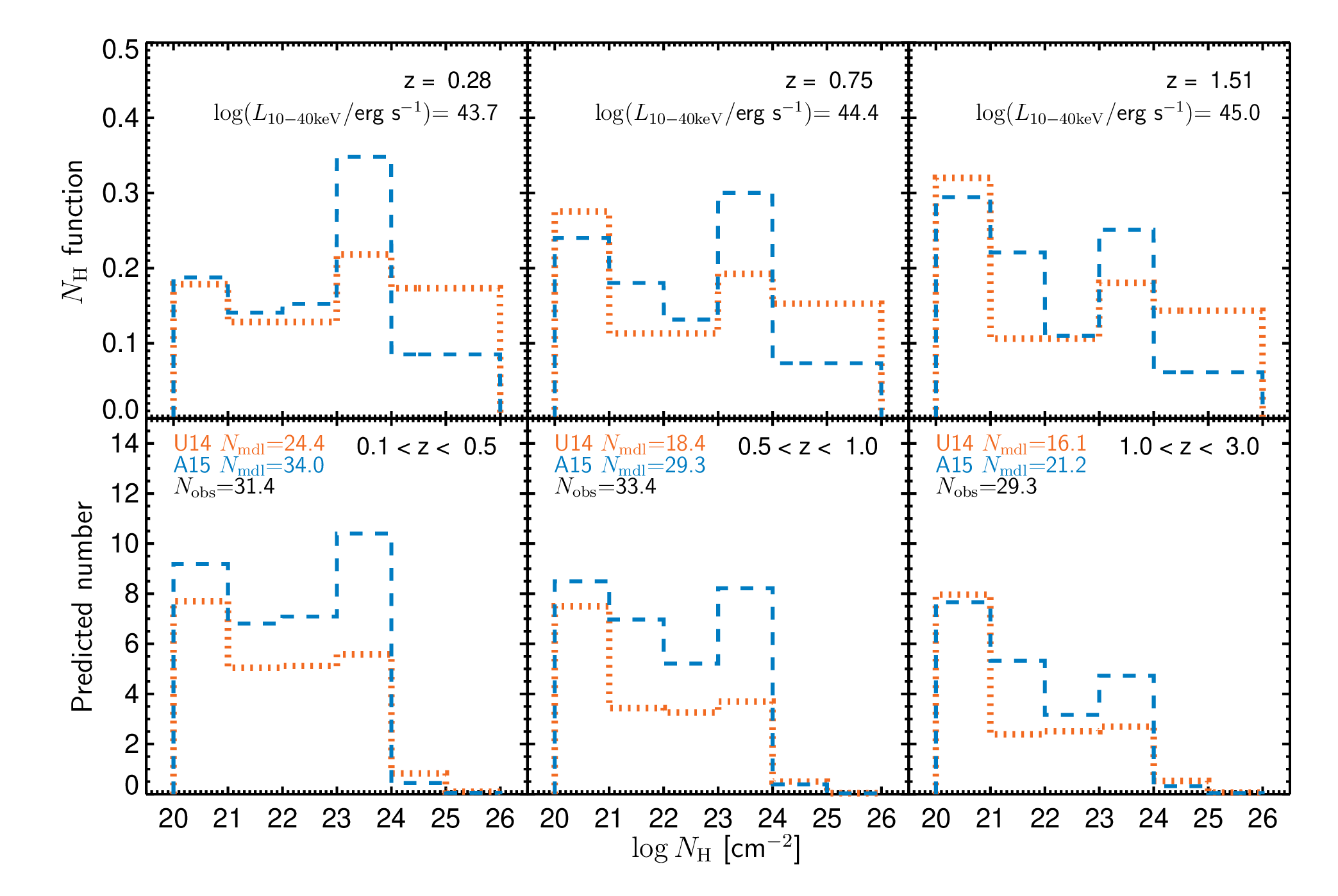}
\caption{
Model predictions of the intrinsic and observable \NH\ functions.
\emph{Top panels:} Intrinsic \NH\ functions evaluated at the mean redshift and geometric mean luminosity of sources in the bin, based on the A15 (blue dashed line) and U14 (red dotted line) models.
\emph{Bottom panels:}
Predicted numbers of sources in our \emph{NuSTAR} 8--24~keV selected sample in each redshift range as a function of \NH, based on the A15 and U14 models \refone{(folded through our sensitivity and assuming our baseline $R=1$ spectral model in both cases)}. 
We also give the total predicted numbers ($N_\mathrm{mdl}$) in each redshift range and the observed number of sources ($N_\mathrm{obs}$). The total observed number in a bin ($N_\mathrm{obs}$) is non-integer due to the small number of sources with undetermined redshifts that can thus make a partial contribution to multiple redshift bins.
}
\label{fig:nhdist}
\end{figure*}

To explore this further, in the top panels of Figure~\ref{fig:nhdist} we show the \emph{intrinsic} \NH\ functions from the U14 and A15 models, evaluated at the mean redshift and geometric mean luminosity for \emph{NuSTAR} sources in each bin. 
There are clear differences whereby the U14 model has a higher fraction of sources in the Compton-thick regime ($\fcthick=50$\% versus 25\% for A15) and a lower relative fraction of sources with $\nh=10^{23-24}$ \cmsq\ (i.e. heavily obscured, but Compton-thin sources). 
These differences result in higher binned estimates of the XLF in Figure~\ref{fig:xlfbins} when the U14 model is used to correct for selection biases -- the binned estimates are increased to allow for a larger population of Compton-thick sources that are not represented in our observed samples. 

The bottom panels of Figure~\ref{fig:nhdist} show the predicted numbers of sources with different \NH\ values in our 8--24 keV selected sample.
These are calculated by folding the U14 or A15 models of the \NH\ function (and XLF) through our sensitivity functions (assuming our X-ray spectral model) over the entire redshift bin.
The predicted distributions differ substantially from the intrinsic \NH\ functions shown in the top panels; most noticeably the predicted numbers of Compton-thick AGNs in the samples are extremely small.
\refone{Such small numbers are due to the flux from a Compton-thick AGN being suppressed by a factor of $\gtrsim2-10$, even in the relatively hard 8--24~keV band that is used for selection, combined with the steep slope of the XLF at the bright luminosities that we probe, which results in strong selection biases against such sources} \citep[\refone{see Figure~\ref{fig:lx_vs_nh}} and][]{Ballantyne11}.
The predicted numbers of sources based on the U14 model are generally smaller than with the A15 model due to the higher intrinsic $\fcthick$ in the U14 model. 
The total predicted numbers, $N_\mathrm{mdl}$ combined over all luminosities, and the effective observed numbers in a bin, $N_\mathrm{obs}$, are given in each panel.
Generally, $N_\mathrm{mdl}$ based on the A15 model is closer to $N_\mathrm{obs}$ in all of the redshift panels, although statistically the $N_\mathrm{mdl}$ estimates from both A15 and U14 are consistent with $N_\mathrm{obs}$ at the $<2.5\sigma$ level. 
Combining the three redshift panels, we find that our total observed number of sources (94) is consistent with the prediction based on A15 (84.5) at the $<1\sigma$ level but is significantly higher ($>3\sigma$) than the U14 prediction (58.9).

Another possible explanation for the U14 model under-predicting the observed number of sources could be our choice of X-ray spectral model. 
Assuming different prior distributions for $\Gamma$, $f_\mathrm{scatt}$ and $R$ could change 
the extrapolation of our models of the XLF to the 10--40 keV band, potentially bringing the model XLF into better agreement with our binned estimates and altering our $N_\mathrm{mdl}$ predictions (without requiring any changes to the \NH\ function).
Varying $f_\mathrm{scatt}$ has a minimal impact as the observed \mbox{8--24~keV} band probes the direct emission and reflection component, rather than the scattered emission that is important at lower energies. Varying $\Gamma$ has a larger impact, although a very flat, intrinsic photon index ($\Gamma\approx1.4$ which is then subjected to absorption) is required to bring the U14 model into agreement with our binned estimates.
It is now well established that the intrinsic photon index of the power-law emission from AGNs has a mean $\Gamma\approx1.9$ with scatter $\lesssim0.2$ \citep[e.g.][]{Nandra94,Tozzi06,Scott11}, as assumed in our original spectral model.

Changing our assumed reflection strength, $R$, has a more significant impact. 
Our baseline spectral model assumes a flat distribution in the range $R=0-2$ and thus an average reflection of $\langle R \rangle=1$.
Assuming stronger reflection increases the luminosity in the 10--40 keV band for the same 2--10 keV luminosity.\footnote{The contribution from reflection is negligible at the $\sim$2--10~keV energies probed by A15 and U14, thus adopting a spectral model with stronger reflection when extrapolating to the \emph{NuSTAR} band does not contradict the results of these lower energy studies.} For $R=2$, we find 
\begin{equation}
\log (\lnu / \mathrm{erg\;s^{-1}}) \approx \log (L_\mathrm{2-10\;keV} / \mathrm{erg\;s^{-1}}) + 0.26
\label{eq:ref2}
\end{equation}
with our spectral model,
which shifts the model XLF to higher \Lnu\ (grey dotted line in Figure~\ref{fig:xlfbins}) and brings it into good agreement with the U14-based binned estimates.
For a fixed $R=2$ (for all sources), we predict $N_\mathrm{mdl}=99.9$ (across all redshift bins) with the U14 model of the \NH\ function, in good agreement with our observed number of sources (94). 
However, with the A15 model of the \NH\ function we predict $N_\mathrm{mdl}=146.2$ if we adopt a spectral model with stronger reflection ($R=2$), thus over-predicting the observed number of sources.

In conclusion, the U14 model of the \NH\ function predicts \emph{fewer} sources than observed in our \mbox{8--24~keV} sample, possibly due to a higher relative fraction of Compton-thick AGNs (compared to A15). 
Thus, the binned estimates of the XLF are slightly higher.
The A15 model, which predicts more sources with $\nh\approx10^{23-24}$~\cmsq, is in better agreement with the observed samples. 
Alternatively, a stronger reflection component (average $R\approx2$ rather than $R\approx1$) across all luminosities results in better agreement between our measurements of the XLF and the extrapolation of the U14 model. 
Further study (e.g. X-ray spectral analysis) is required to improve constraints on the \NH\ function, determine the intrinsic distribution of $R$, and refine our estimates of the XLF.

\begin{deluxetable*}{llllll}
\tablecaption{Binned estimates of the rest-frame 10--40 keV X-ray luminosity function of AGNs based on the \emph{NuSTAR} 8--24 keV sample.
\label{tab:xlf}}
\tablehead{
\colhead{$z$-range} & \colhead{$\langle z \rangle$\tablenotemark{a}} & \colhead{$\log L_\mathrm{10-40keV}$} &
\colhead{$\phi_\mathrm{noabs}$\tablenotemark{b}} & \colhead{$\phi_\mathrm{A15}$\tablenotemark{c}} & \colhead{$\phi_\mathrm{U14}$\tablenotemark{d}}\\
                    &                                                &  \colhead{ [\ergs] }               &
\colhead{[Mpc$^{-3}$ dex$^{-1}$]} & \colhead{[Mpc$^{-3}$ dex$^{-1}$]} & \colhead{[Mpc$^{-3}$ dex$^{-1}$]} 
}
\startdata
  0.1--0.5 & 0.28  &   42.75--43.00 & $1.81^{+2.14}_{-1.08} \times 10^{-4}$ & $2.64^{+3.86}_{-1.76} \times 10^{-4}$ & $3.42^{+4.87}_{-2.25} \times 10^{-4}$ \\
           &       &   43.00--43.25 & $9.39^{+8.13}_{-4.67} \times 10^{-5}$ & $1.52^{+1.46}_{-0.81} \times 10^{-4}$ & $1.74^{+1.71}_{-0.93} \times 10^{-4}$ \\
           &       &   43.25--43.50 & $5.69^{+3.50}_{-2.27} \times 10^{-5}$ & $9.51^{+5.91}_{-3.82} \times 10^{-5}$ & $1.02^{+0.66}_{-0.42} \times 10^{-4}$ \\
           &       &   43.50--43.75 & $2.63^{+1.46}_{-0.98} \times 10^{-5}$ & $4.08^{+2.28}_{-1.52} \times 10^{-5}$ & $4.60^{+2.62}_{-1.74} \times 10^{-5}$ \\
           &       &   43.75--44.00 & $8.48^{+6.33}_{-3.85} \times 10^{-6}$ & $1.27^{+0.86}_{-0.54} \times 10^{-5}$ & $1.59^{+1.07}_{-0.67} \times 10^{-5}$ \\
           &       &   44.00--44.25 & $3.20^{+2.95}_{-1.66} \times 10^{-6}$ & $4.28^{+3.84}_{-2.18} \times 10^{-6}$ & $5.73^{+5.03}_{-2.87} \times 10^{-6}$ \\
           &       &   44.25--44.50 & $1.81^{+1.71}_{-0.95} \times 10^{-6}$ & $3.00^{+2.63}_{-1.50} \times 10^{-6}$ & $3.83^{+3.28}_{-1.90} \times 10^{-6}$ \vspace{5pt}\\
\tableline\\                                                              $   $
  0.5--1.0 & 0.75  &   43.50--43.75 & $1.74^{+3.09}_{-1.27} \times 10^{-4}$ & $2.12^{+4.89}_{-1.73} \times 10^{-4}$ & $2.95^{+7.05}_{-2.44} \times 10^{-4}$ \\
           &       &   43.75--44.00 & $5.92^{+5.42}_{-3.05} \times 10^{-5}$ & $9.49^{+9.21}_{-5.06} \times 10^{-5}$ & $1.25^{+1.20}_{-0.66} \times 10^{-4}$ \\
           &       &   44.00--44.25 & $2.32^{+1.50}_{-0.96} \times 10^{-5}$ & $3.21^{+2.08}_{-1.33} \times 10^{-5}$ & $3.89^{+2.48}_{-1.59} \times 10^{-5}$ \\
           &       &   44.25--44.50 & $1.53^{+0.57}_{-0.42} \times 10^{-5}$ & $1.99^{+0.76}_{-0.56} \times 10^{-5}$ & $2.26^{+0.86}_{-0.64} \times 10^{-5}$ \\
           &       &   44.50--44.75 & $4.17^{+1.98}_{-1.39} \times 10^{-6}$ & $5.89^{+2.65}_{-1.89} \times 10^{-6}$ & $6.42^{+2.94}_{-2.08} \times 10^{-6}$ \\
           &       &   44.75--45.00 & $3.74^{+6.38}_{-2.68} \times 10^{-7}$ & $5.50^{+7.62}_{-3.57} \times 10^{-7}$ & $5.84^{+8.49}_{-3.88} \times 10^{-7}$ \vspace{5pt}\\
\tableline\\                                                              $   $
  1.0--3.0 & 1.51  &   44.25--44.50 & $4.21^{+4.73}_{-2.44} \times 10^{-5}$ & $5.75^{+7.36}_{-3.58} \times 10^{-5}$ & $1.03^{+1.28}_{-0.63} \times 10^{-4}$ \\
           &       &   44.50--44.75 & $1.61^{+1.16}_{-0.71} \times 10^{-5}$ & $2.08^{+1.57}_{-0.95} \times 10^{-5}$ & $3.34^{+2.56}_{-1.54} \times 10^{-5}$ \\
           &       &   44.75--45.00 & $5.69^{+3.09}_{-2.09} \times 10^{-6}$ & $6.75^{+3.70}_{-2.49} \times 10^{-6}$ & $8.92^{+4.95}_{-3.32} \times 10^{-6}$ \\
           &       &   45.00--45.25 & $2.00^{+1.00}_{-0.69} \times 10^{-6}$ & $2.36^{+1.17}_{-0.81} \times 10^{-6}$ & $2.68^{+1.33}_{-0.92} \times 10^{-6}$ \\
           &       &   45.25--45.50 & $3.76^{+3.19}_{-1.85} \times 10^{-7}$ & $4.86^{+3.81}_{-2.27} \times 10^{-7}$ & $4.89^{+3.84}_{-2.29} \times 10^{-7}$ \\
           &       &   45.50--45.75 & $1.56^{+1.76}_{-0.91} \times 10^{-7}$ & $1.54^{+1.66}_{-0.87} \times 10^{-7}$ & $1.40^{+1.51}_{-0.79} \times 10^{-7}$ \vspace{5pt}
\enddata
\tablenotetext{a}{Mean redshift of sources in the redshift bin.}
\tablenotetext{b}{Binned estimates of the XLF, neglecting absorption (i.e. assuming all sources have $\nh=10^{20}$ \cmsq). The A15 model for the XLF of Compton-thin AGNs is adopted for the $N_\mathrm{obs}/N_\mathrm{mdl}$ method. Errors are based on the Poisson error in the observed number of sources ($N_\mathrm{obs}$).}
\tablenotetext{c}{Binned estimates of the XLF, adopting the A15 model for the \NH\ function and the total XLF (including Compton-thick sources) for the $N_\mathrm{obs}/N_\mathrm{mdl}$ method.}
\tablenotetext{d}{Binned estimates of the XLF, adopting the U14 model for the \NH\ function and the total XLF (including Compton-thick sources) for the $N_\mathrm{obs}/N_\mathrm{mdl}$ method.}
\end{deluxetable*}

\section{Discussion}
\label{sec:discuss}

\subsection{The evolution of the XLF}

\begin{figure}
\includegraphics[trim=10 10 0 0,width=\columnwidth]{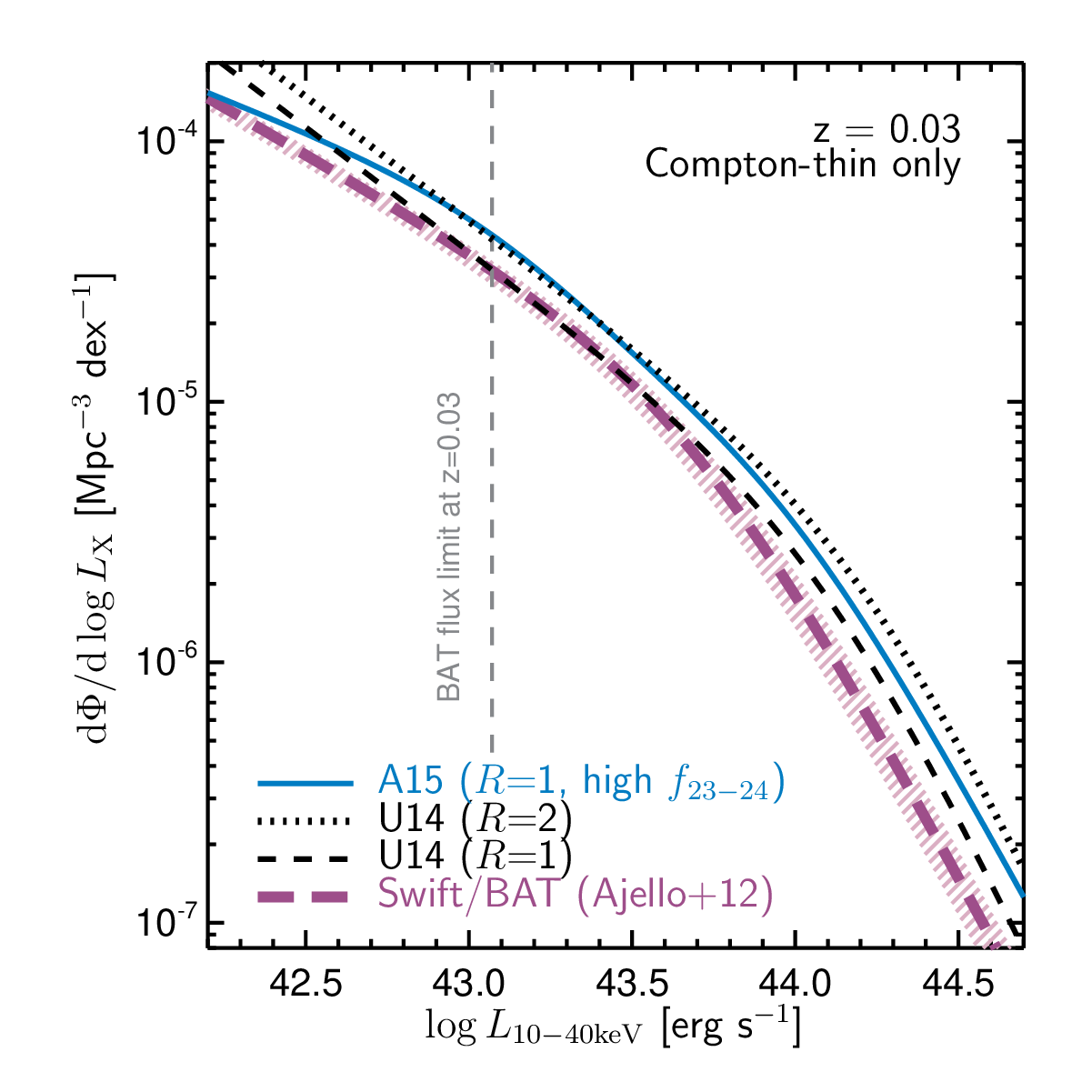}
\caption{
\refone{
Comparison of the local XLF of Compton-thin AGNs, based on the best fit to the \textit{Swift}/BAT 60-month 15--55~keV sample by \citet{Ajello12} (purple long-dashed line, hatched region indicates conversion to 10--40~keV with a photon-index between $\Gamma=1.4$ and $\Gamma=2.1$) 
to extrapolations of the U14 and A15 models (evaluated at $z=0.03$, the median redshift of the \emph{Swift}/BAT sample). 
The extrapolations of the A15 model with moderate reflection ($R=1$) and a relatively high fraction of sources with $\nh=10^{23-24}$~\cmsq\ (blue solid line) and the U14 model with strong reflection ($R=2$, dotted black line), both of which can describe the high-redshift population seen by \emph{NuSTAR}, over-predict the observed $z\sim0$ XLF at all luminosities. 
The U14 model with $R=1$ is in better agreement with the \emph{Swift}/BAT measurement close to the flux limit (vertical grey dashed line); however, this model significantly under-predicts the number of sources in our high-redshift \emph{NuSTAR} sample. 
These findings indicate that there is evolution in the XLF, absorption distribution, or spectral properties of AGNs between $z\sim0$ and the higher redshifts ($z\sim0.1-3$) probed by {\emph NuSTAR} that is not accounted for by the current models.
}
}
\label{fig:xlf_z0}
\end{figure}

We have presented the first measurements of the rest-frame 10--40 keV XLF of AGNs at high redshifts ($0.1<z<3$) based on a sample of sources selected at comparable observed-frame energies (8--24 keV). 
Selecting at hard X-ray energies---in contrast to prior studies based on lower energy data---provides estimates of the intrinsic X-ray luminosity that are relatively unaffected by absorption due to Compton-thin column densities (\mbox{$\nh\lesssim10^{24}$~\cmsq}). 
Thus, our work provides the most direct measurements of the XLF currently possible at these redshifts, although our selection remains biased against Compton-thick sources. 

Our results \refone{are consistent} with the strong evolution of the AGN population seen in lower energy studies of the XLF \citep[e.g.][]{Ueda03,Barger05} and at longer wavelengths \citep[e.g.][]{Ross12,Lacy15},
characterized by a shift in the luminosity function toward higher luminosities at higher redshifts.
\refone{This evolution leads to the substantial increase in the space density of luminous ($\lnu \gtrsim 10^{44}$ \ergs) AGNs out to $z\sim 2$ that is seen in our measurements of the XLF (see Figure \ref{fig:xlfbins}).}
 \refone{
No substantial modifications to this overall picture appear to be required based on our \emph{NuSTAR} data, although the limited range of luminosities probed by \emph{NuSTAR} means we are unable to measure the faint-end slope of the XLF at $z\gtrsim0.5$ or compare different parameterizations.
Furthermore, we find that strong ($R\sim2$) reflection is needed to reconcile the U14 model of the XLF with our \emph{NuSTAR} sample (see further discussion in Section~\ref{sec:degeneracy} below).}

\refone{Our \emph{NuSTAR} measurements provide a high-redshift comparison to previous measurements of the XLF at hard ($\gtrsim10$~keV) X-ray energies, which have been restricted to the local ($z\lesssim0.1$) Universe. 
In Figure~\ref{fig:xlf_z0} we compare the best fit to the XLF of local AGNs in the \emph{Swift}/BAT 60-month sample \citep[thick purple dashed line,][]{Ajello12} with extrapolations of the U14 and A15 models to $z=0.03$ (the median redshift of sources in the \emph{Swift}/BAT sample). 
Both the A15 model (solid blue line)
and the U14 model with a strong ($R=2$) reflection component (black dotted line)---either of which is consistent with our higher redshift \emph{NuSTAR} data---\emph{over}-predict the local \emph{Swift}/BAT XLF at all luminosities.
The U14 model with $R=1$ is in better agreement with the \emph{Swift}/BAT XLF close to the flux limit (vertical grey dashed line); however, this model significantly \emph{under}-predicts the number of sources in our high-redshift \emph{NuSTAR} sample (see Figure~\ref{fig:nhdist}, discussed in Section~\ref{sec:xlf} above).
A similar result is seen in our analysis of the \emph{NuSTAR} number counts, where the predictions of population synthesis models (based on either the U14 XLF with strong reflection or the A15 model) provide a good agreement with the observed \emph{NuSTAR} number counts but over-predict the counts at the brighter fluxes probed by \emph{Swift}/BAT
\citep[see figure 5 of][]{Harrison15}.
\citet{Ballantyne14} also highlighted the discrepancies between measurements of the local XLF and the extrapolations of high-redshift evolutionary models. 
These findings indicate that there must be some evolution in the XLF, absorption distribution, or spectral properties of AGNs between $z\sim0$ and the higher redshifts probed by \emph{NuSTAR} that is not fully accounted for by the current models.
}

\subsection{The absorption distribution, the fraction of Compton-thick sources and the reflection strength}
\label{sec:degeneracy}

Our final binned estimates of the XLF differ slightly depending on the underlying model of the \NH\ function that is used to correct for selection biases, whereby the binned estimates using the U14 model are systematically higher than the binned estimates using the A15 model.
Our observed sample appears to favor a higher relative number of heavily absorbed ($\nh\approx10^{23-24}$ \cmsq) sources and a lower Compton-thick fraction, as given by the A15 model.
We note that the \NH\ function of A15 is based on an indirect method, attempting to reconcile samples of AGNs selected at \mbox{0.5--2~keV} and \mbox{2--7~keV} energies from \emph{Chandra} surveys, spanning out to $z\sim5$. 
In contrast, U14 determine the \NH\ function in the local Universe based on spectral analysis of sources in the \emph{Swift}/BAT AGN sample (selected at 15--195~keV), which is then extrapolated to higher redshifts.
Both of these methods have their limitations. 
Accurate constraints on the intrinsic \NH\ function in both the local and high-redshift universe are vital to determine the extent of obscured black hole growth and shed light on the physical origin of the obscuring material \citep[e.g.][]{Hopkins06b,Buchner15}.
While our results provide indirect constraints on the \NH\ function (we do not measure \NH\ for individual sources), our measurements of the 10--40~keV XLF place 
important constraints on the number densities of absorbed, moderately luminous sources at $z\gtrsim0.1$.

The incidence and strength of Compton reflection, which can substantially boost the observed fluxes at $\gtrsim10$~keV energies, also has an important impact on our study.
Moderate reflection ($R\approx1$) provides a good agreement between our observed \emph{NuSTAR} sample and the extrapolation of the A15 XLF from lower energies.
A stronger ($R\approx2$) reflection component is needed to
reconcile our \emph{NuSTAR} sample with the extrapolated model without requiring changes to the U14 \NH\ function and Compton-thick fraction. 
Most prior studies have found that $R$ (usually probed indirectly via the strength of the iron K-line) is inversely correlated with luminosity and generally weak (i.e. $R\lesssim 1$) for moderately luminous AGNs \citep[e.g.][]{Iwasawa93,Nandra97b,Page05,Ricci11}.
Spectral analysis of a single source in our \emph{NuSTAR} sample 
\citep[NuSTAR J033202--2746.8 in the ECDFS:][]{DelMoro14} found a moderate reflection component ($R\approx0.55$) for this high luminosity ($\lnu\approx6.4 \times 10^{44}$ \ergs) source. 
However, \citet{Ballantyne14} found that strong reflection ($R\approx1.7$) at all luminosities is needed to reconcile different measurements of the local XLF across a wide range of X-ray energies ($\sim$0.5--200~keV).
Our measurements of the 10--40~keV XLF indicate that moderate-to-strong reflection ($R\sim1-2$) is required to describe the average spectral characteristics of $\lnu\sim10^{43-46}$~\ergs\ AGNs at $z\sim 0.1-3$.
The extent, strength and spectral characteristics of reflection provides insights into the physical nature of the obscuring material and the accretion disk \citep[e.g.][]{Garcia13,Falocco14,Brightman15}.
Strong reflection could also indicate a substantial population of rapidly spinning black holes in the detected sample; however, a relatively small intrinsic fraction of high-spin sources ($\sim7$\%) can potentially dominate the observed number counts at a given flux limit \citep[][]{Brenneman11,Vasudevan15}.
Accurately measuring the distribution of reflection is thus an important challenge for future statistical studies of AGN populations.

The Compton-thick fraction and the strength of reflection are also vital parameters for understanding the origin of the CXB, in particular, the peak at $\sim$20--30~keV \citep[e.g.][]{Gilli07,Draper09,Akylas12}.
The degeneracy between these parameters and the consequences for AGN population synthesis models are discussed in detail by \citet{Treister09}.
Our measurements of the XLF appear to suffer from a similar degeneracy.
Nevertheless, our results constrain the distribution of luminosity and redshift of sources 
responsible for a large fraction of the CXB emission \citep[$\sim35$\% at 8--24~keV, see][]{Harrison15}, placing an additional constraint on the possible contributions of Compton-thick AGNs or reflection to the CXB.

Recent \emph{NuSTAR} studies 
have 
provided constraints on the distribution of \NH\ of optically selected Type-2 QSOs \citep{Gandhi14,Lansbury14,Lansbury15} and find that \NH\ is often underestimated for these sources based purely on low energy ($<10$~keV) data.
Given the small sample sizes and large remaining uncertainties, their recovered \NH\ function is consistent with both the U14 and A15 models adopted in this paper. 
C15 use a band ratio analysis to identify candidate Compton-thick AGNs among \emph{NuSTAR} detected sources in the survey of the COSMOS field.
Their number of candidates ($\sim13-20$\% of the \emph{NuSTAR} sources) is significantly higher than our predictions 
(we expect only $\sim0.5$ Compton-thick AGNs out of 32 sources detected at 8--24~keV energies in COSMOS), indicating that our models may need updating to a much higher intrinsic $\fcthick$. 
However, X-ray spectral analysis is required to measure \NH\ and confirm the Compton-thick nature of these candidates.
\citet{Alexander13} presented X-ray spectral analysis of the first ten sources from the \emph{NuSTAR} serendipitous survey, finding a high fraction ($\gtrsim50$\%) have $\nh\gtrsim10^{22}$~\cmsq\ but none was Compton-thick, consistent with the predictions of our models, albeit for a limited sample size. 
\citet{DelMoro14} presented the first spectral analysis from the deep survey of the ECDFS, focusing on a single source.

\subsection{Future prospects}

To fully test the conclusions of this paper, compare between the U14 and A15 models, and refine our estimates of the XLF requires accurate measurements of the distribution of \NH\ and $R$ for AGNs across a wide range of luminosities and redshifts.
Spectral analysis of sources from across the \emph{NuSTAR} extragalactic survey program will be the focus of forthcoming papers (Zappacosta et~al. in preparation, Del~Moro et~al. in preparation) and will place crucial constraints on the distribution of \NH\ and $R$ for luminous AGNs out to $z\sim3$.
Ultimately though, the ability of the \emph{NuSTAR} survey program to constrain the XLF, \NH\ function, Compton-thick fraction, or reflection properties of AGNs is limited by both depth and sample size.
The ongoing survey program will improve this situation.
Our serendipitous sample will roughly double with the inclusion of Southern fields 
(see Lansbury et al. in preparation)
and continues to grow as \emph{NuSTAR} observes targets.
The dedicated survey program is also continuing and will initially focus on increasing the area coverage of the deep ($\sim400$ ks) layer via observations of the GOODS-N \citep{Alexander03} and UDS \citep{Lawrence07,Ueda08} 
fields. 
These observations will increase the area coverage at the faintest fluxes by $\gtrsim50$\%, improving the number statistics at low luminosities.

Pushing to greater depths, however, is vital to constrain the low-to-moderate luminosity AGN population ($\lesssim L_*$) that corresponds to the bulk of the accretion density.
Our current \emph{NuSTAR} sample is limited to luminous X-ray sources.
We do not probe below the break in the XLF ($L_*$) at $z\gtrsim0.5$ and do not place strong constraints on the faint-end slope in any redshift bin.
Thus, we are unable to address issues regarding the best parametric description of the evolution of the XLF (i.e. pure luminosity evolution, independent luminosity and density evolution, luminosity-dependent density evolution) or the extent of any evolution in the shape \citep[see][U14, A15]{Aird10,Miyaji15}. 
\refone{At fainter luminosities the intrinsic Compton-thick fraction is expected to be somewhat higher and the flatter slope of the XLF should reduce the biases against the detection of such sources in the current samples (see Figure~\ref{fig:nhdist}). 
Indeed, the observed fraction of Compton-thick AGNs in deep \textit{Chandra} surveys \citep[$\sim3$\%, e.g.][]{Brightman14} is similar to the expected fraction in our \emph{NuSTAR} survey and the absolute numbers \citep[$\sim100$ Compton-thick candidates were identified by][]{Brightman14} are much larger, mainly due to the much fainter luminosities that are accessed by \textit{Chandra}.}\footnote{\refone{We note that the higher energies probed by \emph{NuSTAR} are vital to accurately characterize the spectra of both Compton-thin and Compton-thick AGNs, measure \NH, and determine the strength of the reflection component \citep[see][]{Lansbury15}.}}
Substantially increasing the nominal depths of the \emph{NuSTAR} survey is challenging as the observations are background limited for exposures $\gtrsim150$~ks.
An alternative strategy is to consider sources that are detected in the broader 3--24~keV band,
providing a sample of sources a factor $\gtrsim2$ larger and reaching luminosities a factor $\sim2$ fainter than the 8--24 keV sample considered here.
A preliminary analysis of the 3--24~keV sample (using the same, indirect approach to absorption corrections described in this paper) indicates that the XLF is consistent with the U14 and A15 models at $\lesssim L_*$, although this band is dominated by soft ($\lesssim 8$~keV) photons and larger, uncertain corrections are required for the fraction of absorbed and Compton-thick sources.
However, many of the sources detected at 3--8~keV or 3--24~keV in the \emph{NuSTAR} surveys have statistically significant counts (and thus useful information) at harder energies.
Thus, it may be possible to improve constraints on the XLF, \NH\ function, and distribution of reflection via sophisticated analysis of the band ratios or X-ray spectra, or via stacking analyses.
Careful consideration of the survey sensitivity and selection biases will be vital in such studies.

\section{Conclusions}
\label{sec:conclusions}

\begin{itemize}[leftmargin=*]
\item
We have presented the first measurements of the rest-frame 10--40 keV XLF of AGNs at $0.1<z<3$ based on a sample of 94 sources selected at comparable observed-frame energies (8--24 keV). 
Our study takes advantage of the unprecedented sensitivity at these energies that is achieved by the \emph{NuSTAR} survey program.

\item
We find that different models of the \NH\ function, used to account for selection biases in our measurements, make significantly different predictions for the total number of sources in our sample, leading to slight differences in our binned estimates of the XLF.
The \citet{Ueda14} model predicts fewer AGNs than observed in our sample, possibly due to a higher \Fcthick, whereas the \citet{Aird15} model, which predicts more sources with $\nh\approx10^{23-24}$~cm$^{-2}$, is in better agreement with the observed samples.

\item 
Our results are also sensitive to our assumed X-ray spectral model.
Stronger reflection ($R\approx2$, compared to our baseline assumption of $R\approx1$) at all luminosities can bring the \citet{Ueda14} model predictions into better agreement with our \emph{NuSTAR} sample.
However, with $R\approx2$ the \citet{Aird15} model over-predicts the number of sources in our sample by $\gtrsim50$\%.

\item
Our results \refone{are consistent with} the strong evolution of the AGN population seen in lower energy studies of the XLF \citep[e.g.][]{Ueda03,Barger05,Aird10}, characterized by a shift in the luminosity function toward higher luminosities at higher redshifts. 
\refone{However, the models that successfully describe the high-redshift population detected by \emph{NuSTAR} tend to over-predict the local ($z\approx0$) XLF measured by \emph{Swift}/BAT, indicating some evolution of the AGN population that is not fully captured by the current models.
Nonetheless, as our sample is limited to luminous ($\gtrsim L*$) X-ray sources at $z\gtrsim0.5$,  we defer an investigation of different parametric descriptions of the evolution of the XLF to future studies.
}

\item
Forthcoming X-ray spectral analysis of the \emph{NuSTAR} survey should enable us to measure \NH\ and $R$ for the brightest sources, 
\refone{break the degeneracy between the \NH\ function and the average reflection strength},
and refine our estimates of the XLF.
Including lower-energy \emph{NuSTAR} detections may enable us to probe a factor~$\sim$2 deeper.  
The ongoing \emph{NuSTAR} survey program will also increase our sample size and improve our estimates of the XLF.
\end{itemize}

\acknowledgements
This work made use of data from the \emph{NuSTAR} mission, a project led by the California Institute of Technology, managed by the Jet Propulsion Laboratory, and funded by
the National Aeronautics and Space Administration. We thank the \emph{NuSTAR} Operations, Software and Calibration teams for support with the execution and analysis of these
observations. This research has made use of the \emph{NuSTAR} Data Analysis Software (NUSTARDAS) jointly developed by the ASI Science Data Center (ASDC, Italy) and the California Institute of Technology (USA).
We acknowledge financial support from:
ERC Advanced Grant FEEDBACK at the University of Cambridge (JA, ACF); 
a COFUND Junior Research Fellowship from the Institute of Advanced Study, Durham University (JA);
the  Science  and  Technology  Facilities  Council  (STFC)  grants 
ST/I001573/1 (DMA and ADM),
ST/K501979/1  (GBL), and ST/J003697/1 (PG);
the Leverhulme Trust (DMA);
the Caltech Kingsley Visitor Program (DMA, AC);
NSF award AST 1008067 (DRB);
NASA grants 11-ADAP11-0218 and GO3-14150C (FC);
an Alfred P. Sloan Research Fellowship and a Dartmouth Class of 1962 Faculty Fellowship (RCH);
CONICYT-Chile grants Basal-CATA PFB-06/2007 (FEB, ET);
FONDECYT 1141218 (FEB) and 1120061 (ET);
``EMBIGGEN" Anillo ACT1101 (FEB, ET); 
the Ministry of Economy, Development, and Tourism’s Millennium Science Initiative through grant IC120009, awarded to The Millennium Institute of Astrophysics, MAS (FEB);
NASA NuSTAR subcontract 44A-1092750 and NASA ADP grant NNX10AC99G (WNB, BL);
ASI/INAF grant I/037/12/0011/13 (AC, LZ);
and NASA Earth and Space Science Fellowship Program grant NNX14AQ07H (MB).

\end{document}